\documentclass[journal,10pt]{IEEEtran}
\pdfoutput=1

\usepackage{cite}

\usepackage{amsmath}
\usepackage{algorithm}
\usepackage{algorithmic}
\usepackage{subfigure}
\usepackage{amssymb}
\usepackage{graphicx}
\usepackage{verbatim}
\usepackage{empheq}

\usepackage{array}
\usepackage{mathtools}
\usepackage{graphicx}
\usepackage{cases}
\usepackage[colorinlistoftodos]{todonotes}
\usepackage[colorlinks=true, linkcolor=black,citecolor=black,urlcolor=black]{hyperref}
\usepackage{indentfirst}
\usepackage{multirow}
\usepackage{booktabs}
\usepackage{makecell}

\usepackage[switch]{lineno}

\allowdisplaybreaks[4]

\begin{document}


\title{Energy and Age Pareto Optimal Trajectories in UAV-assisted Wireless Data Collection}
\author{
Yuan~Liao, \IEEEmembership{Student Member,~IEEE}  and Vasilis~Friderikos, \IEEEmembership{Member,~IEEE} 

\thanks{

The authors are with the Department of Engineering, King's College London, London WC2R 2LS, U.K (e-mail: yuan.liao@kcl.ac.uk; vasilis.friderikos@kcl.ac.uk).}
}


\maketitle

\begin{abstract}
This paper studies an unmanned aerial vehicle (UAV)-assisted wireless network, where a UAV is dispatched to gather information from ground sensor nodes (SN) and transfer the collected data to a depot. The information freshness is captured by the age of information (AoI) metric, whilst the energy consumption of the UAV is seen as another performance criterion. Most importantly, the AoI and energy efficiency are inherently competing metrics, since decreasing the AoI requires the UAV returning to the depot more frequently, leading to a higher energy consumption. To this end, we  design UAV paths that optimize these two competing metrics jointly and reveal the Pareto frontier. To formulate this problem,  a multi-objective mixed integer linear programming (MILP) is proposed with a flow-based constraint set and we apply Bender's decomposition on the proposed formulation. Numerical results show that the proposed method allows deriving non-dominated solutions among two competing metrics when designing the UAV path.
\end{abstract}


\IEEEpeerreviewmaketitle

\section{Introduction}
\label{introduction}

\begin{figure}[!t]
\centering
\subfigure{\includegraphics[width=.38\textwidth]{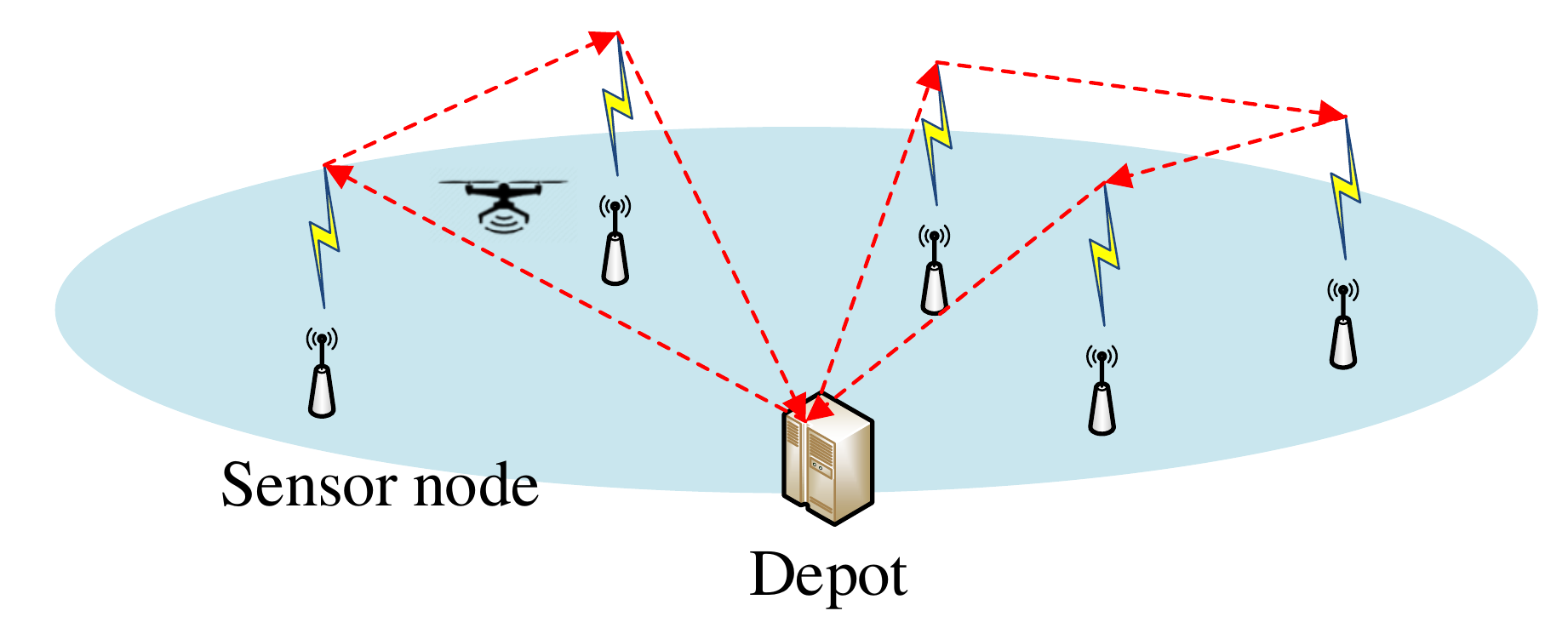}}
\caption{Illustration of the proposed multi-return-allowed mode}
\label{ToyExample}
\end{figure}

\IEEEPARstart{D}ue to their high flexibility and controllable mobility, one attractive application of unmanned aerial vehicles (UAV) is to serve as data collectors in a wireless sensor network (WSN). To make real-time decisions and enable seamless operation in such networks, the freshness of received information, which is measured by the so-called age of information (AoI), is of critical significance and has received significant attention recently \cite{9380899}. Besides, due to the limited capacity of the on-board batteries, the aspect of energy efficiency is seen as a key challenge in UAV-enabled wireless networks. Revealing the underlying trade-off between those two inherently competing metrics is the main focus in this paper. The two extreme points in the Pareto curve is when only the AoI is minimized and when only the energy consumption is taken into account. In the first case, the UAV creates a star trajectory meaning that to minimize the AoI the UAV returns to the depot every time data are collected from a ground sensor node (SN). On the other extreme, the trajectory will be a Hamiltonian path with the minimal energy consumption. Between those two extreme scenarios are the cases where the UAV returns to the depot after serving a subset of SNs, as shown in Fig.\ref{ToyExample}. The focus on this work is to reveal the non-dominated operating points in the continuum between those two extreme points of operation.  

Several existing works concentrated on the information freshness, i.e., AoI, in UAV-aided wireless networks. In \cite{liu2018age,liu2020uav}, a UAV is applied to collect data from a group of SNs and then return to the depot for further processing with the aim of minimizing the AoI. The following work extends this model to a multi-UAV use case in \cite{mao2020multi}, whereas \cite{wang2017taking} provides overarching architectural aspects of networked operation and connection of multiple UAVs. In \cite{abd2018average}, the UAV is deployed as a relay between a source node and a destination node, and the focus is to provide efficient trajectory design in order to minimize the peak AoI. An AoI-aware UAV trajectory is designed through reinforcement learning based approach in \cite{abd2019deep}. Furthermore, the UAV is applied to sense information from target area directly and the trade-off between the sensing and communication is investigated in \cite{zhang2020age}.

In parallel, several existing research works investigated the energy-efficient path planning since the on-board battery limitation is one of the key challenges in UAV-assisted networks. In \cite{mozaffari2016mobile,zhan2017energy}, the UAV is deployed to collect data from moving ground devices and static SNs, respectively, in which the energy is minimized by trajectory design. A UAV is deployed to cover the most users with minimum transmit power in \cite{alzenad20173}. The energy-efficient trajectory of a rotary-wing UAV is designed with a guaranteed communication requirement in \cite{zeng2019energy}. Furthermore, a closely relevant problem of energy trade-off between the UAV and ground terminals is studied in \cite{yang2018energy}.

Considering the aforementioned competing nature between energy consumption and AoI, we propose a multi-return-allowed serving mode to capture the balance between them, in which the UAV is allowed to return to the depot at any time during the serving cycle. To the best of our knowledge, this is the first work to exploit the Pareto optimality between energy and AoI and create non-dominated decision making on the trajectory planning. Furthermore, although the AoI-aware path has been considered in \cite{liu2018age,mao2020multi,liu2020uav}, these works considered only Hamiltonian paths whereas this work allows several returns to the depot during a serving cycle. Therefore, these previous works can be considered as a special case of our proposed framework. Moreover, to obtain the UAV visiting order initially, we assume that the UAV can only collect data when hovering and propose a multi-objective mixed integer linear programming (MILP) formulation based on a flow-based constraint set, which is solved by Bender's decomposition in a decentralized manner. Afterwards, a more general trajectory in which the UAV can communicate while flying is studied to further improve both metrics. Numerical investigations show that the multi-return-allowed mode achieves the Pareto optimal trade-off between the two competing metrics.

\section{System Model and problem formulation}
\label{systemmodel}

Hereafter, we consider a wireless sensor network with $K$ SNs denoted by the set $\mathcal{K} = \{1,2,...,K\}$, the Cartesian coordinates of which are known and fixed at $\mathbf{w}_i \in \mathbb{R}^2, \, \forall i \in \mathcal{K}$. A rotary-wing UAV acts as a collector to gather information from all SNs and subsequently transmit/offload the data to the depot that located at $\mathbf{w}_0 \in \mathbb{R}^2$. For notational convenience, we combine the depot and SNs to form an extended set $\mathcal{K}^{a} = \{0\} \bigcup \mathcal{K}$, in which the depot is indexed by $0$. 

\subsection{System Model and Age of Information (AoI)}
\label{CommEnermodel}

To obtain the UAV visiting order first, the fly-hover-communication protocol proposed by \cite{zeng2019energy} is adopted in sections \ref{systemmodel} and \ref{LCandBD} for the communication between UAV and SNs, in which the UAV gather information only when hovering right above SNs. Afterwards, a more general solution in which the UAV can communicate while flying is studied in section \ref{FurtherOpt}. Similar to \cite{liu2018age,abd2018average,abd2019deep}, the UAV is assumed to fly at a constant altitude $H$ corresponding to authority regulations and safety considerations,\footnote{In the free-space channel model \eqref{AchRate}, it can be seen that $R$ is monotonically decreasing with $H$. Thus, the UAV would fly at the lowest height corresponding to regulations and safety considerations to achieve the best $R$.} and the achievable rate $R$ for the communication between UAV and SNs is assumed to follow the free-space path loss model as,
\begin{equation}
\label{AchRate}
\begin{aligned}
R = B \cdot \log_2 \Big( 1 + \frac{ P^{t} \rho_0}{\sigma^2H^2 }\Big), 
\end{aligned} 
\end{equation}
where $B$ is the available bandwidth, $P^{t}$ is the transmission power of SNs, $\rho_0$ is the reference channel power gain at $1m$ distance and $\sigma^2$ is the power of channel noise. Furthermore, the hovering duration when the UAV collects data from SN $i$ can be calculated by $T^h_i = D_i / R$, where $D_i$ is the data size to be uploaded from SN $i$. Accordingly, the energy consumption of the UAV when hovering and gathering data from SN $i$ can be calculated by $E^h_i = P^{h} T^h_i$, where $ P^{h}$ is the hovering power of the UAV.

Similar to \cite{liu2018age,mao2020multi,liu2020uav,abd2018average}, the UAV is assumed to keep a fixed speed $V$ so that the flying time between two adjacent SNs can be calculated by $T^f_{ij} = \lVert\mathbf{w}_i - \mathbf{w}_j\rVert / V$, where $\lVert\cdot\rVert$ denotes the 2-norm for a vector \footnote{Initially, the UAV speed is assumed as a constant to better focus on the visiting order, whilst it can be further optimized as the case in section \ref{FurtherOpt}.}. The corresponding energy consumption is $E^f_{ij} = P^{f} T^f_{ij}$, where $P^{f}$ is the propulsion power of the UAV and given by \cite{zeng2019energy} as a function of $V$,
\begin{small}
\begin{equation}
\label{flyingpower}
\begin{aligned}
P^{f}(V) = & P_0\Big(1+\frac{3V^2}{U_{tip}^2} \Big) + P_i \Big(\sqrt{1+\frac{V^4}{4v_0^4}} - \frac{V^2}{2v_0^2} \Big)^{1/2}
\! + \frac{1}{2}d_0\rho sAV^3
\end{aligned} 
\end{equation}
\end{small}where $P_0$ and $P_i$ represent blade profile power and induced power, respectively. $U_{tip}$ is the tip speed of the rotor blade. $v_0$ denotes the mean rotor induced velocity when hovering. $d_0$ and $s$ are the fuselage drag ratio and rotor solidity, respectively. Also, $\rho$ and $A$ denote the air density and rotor disc area, respectively. Subsequently, we define a graph $\mathcal{G} = (\mathcal{K}^{a},\mathcal{E})$, where $\mathcal{E}$ denotes the set of edges, that is, $\mathcal{E} \triangleq \{(i,j)\big| \, \forall i,j  \in \mathcal{K}^{a}, i \neq j\}$. For clarity of exposition, the hovering time and energy consumption at the depot is seen as zero, that is, $T^h_0 = 0$ and $E^h_0 = 0$. Accordingly, we associate the time consumption to the edge $(i,j)$ as,
\begin{equation}
\label{TimeConforEdges}
\begin{aligned}
T_{ij} & =  T^h_i + T^f_{ij} \\ 
& = \begin{cases}
\frac{\lVert\mathbf{w}_i - \mathbf{w}_j\rVert}{V} \qquad \qquad \qquad \qquad \;\;\, \text{ if $i=0, \; (i,j) \in \mathcal{E}$}\\
\frac{D_i}{B \log_2 \Big( 1 + \frac{ P^{t} \rho_0}{\sigma^2H^2 }\Big)} +   \frac{\lVert\mathbf{w}_i - \mathbf{w}_j\rVert}{V} \quad \text{if $i \neq 0, \; (i,j) \in \mathcal{E}$}
\end{cases}
\end{aligned} 
\end{equation}
Similarly, the energy consumption for edge $(i,j)$ is defined as,
\begin{equation}
\label{EnergyConforEdges}
\begin{aligned}
E_{ij} & = E^h_i + E^f_{ij} \\ 
& = \begin{cases}
\frac{P^{f}\lVert\mathbf{w}_i - \mathbf{w}_j\rVert}{V} \qquad \qquad \qquad \qquad \;\;\, \text{ if $i=0, \; (i,j) \in \mathcal{E}$}\\
\frac{P^{h}D_i}{B \log_2 \Big( 1 + \frac{ P^{t} \rho_0}{\sigma^2H^2 }\Big)} +  \frac{P^{f}\lVert\mathbf{w}_i - \mathbf{w}_j\rVert}{V} \quad \text{if $i \neq 0, \; (i,j) \in \mathcal{E}$}
\end{cases}
\end{aligned} 
\end{equation}
Note that both $T_{ij}$ and $E_{ij}$ can be seen as constants once the environment is given. The energy consumption for a given trajectory $\mathcal{Q}$ can be calculated as $ E =  \sum_{(i,j) \in \mathcal{Q}} E_{ij}$, where $(i,j) \in \mathcal{Q}$ means that the UAV travels from $i$ to $j$ adjacently; observe that $E$ is a function of $\mathcal{Q}$.

We adopt the concept of AoI to measure the freshness of information for each SN, and derive the average AoI among all SNs to evaluate the performance of the system. Similar as \cite{liu2018age,mao2020multi,liu2020uav}, the AoI of SN $i$ is defined as the time interval from when the UAV starts collecting information from $i$ to when it returns to the depot. Mathematically, given a cycle consisting of $r$ SNs with the depot as the start (end) vertex, the UAV visits $i_0 \to i_1 \to i_2 ... \to i_r \to i_{r+1}$ in tandem, where $i_0 = i_{r+1} = 0$ and $i_k \in \mathcal{K}, k = 1,2,...,r$. We define the AoI of SNs visited by this cycle recursively, that is,
\begin{linenomath}
\begin{subequations}
\begin{empheq}[left={\empheqlbrace\,}]{align}
& A_{i_r} = T_{i_ri_{r+1}} = T_{i_r0} \label{AoIDef1}\\
& A_{i_{k}} = T_{i_{k}i_{k+1}} + A_{i_{k+1}}, \;\; k = 1,2,...,r-1 \label{AoIDef2}
\end{empheq}
\end{subequations}
\end{linenomath}
where \eqref{AoIDef1} defines the AoI of last visited SN $i_r$ and \eqref{AoIDef2} calculates others.

Furthermore, the average AoI among all SNs, calculated by $\overline{A} = \sum_{i \in \mathcal{K}}A_{i} / K$, is adopted to evaluate the performance of the system. Observe that $\overline{A}$ is a function of the UAV trajectory. But since the trajectory cannot be explicitly expressed, $\overline{A}$ is challenging to be formulated. However, Lemma 3 in \cite{liu2018age} provides an alternative expression of $\overline{A}$ when the trajectory $\mathcal{Q}$ is given, that is,
\begin{equation}
\label{AverageAoI2}
\begin{aligned}
\overline{A} =  \sum_{(i,j) \in \mathcal{Q}}\frac{f_{ij}}{K} T_{ij}
\end{aligned} 
\end{equation}
where $f_{ij}$ denotes the number of visited SNs during the period between the most recent departure from the depot and arrival at $j$. For instance, given a cycle as $i_0 \to i_1 \to i_2 ... \to i_r \to i_{r+1}$ where $i_0 = i_{r+1} = 0$, we have $f_{i_0i_1} = 0, \, f_{i_1i_2} = 1, \,..., \, f_{i_ri_{r+1}} = r$.

\subsection{Flow-based Constraint Set and Problem Formulation}
\label{FBConstraint}

In this subsection, a multi-objective optimization problem is formulated based on a flow-based constraint set \cite{gavish1978travelling} to achieve the trade-off between the average AoI and energy consumption by designing the UAV path.
\begin{linenomath}
\begin{subequations}
\begin{align}
\mathrm{(P1):} 
\quad &  \min_{\mathbf{X}} \sum_{(i,j) \in \mathcal{E}} E_{ij}x_{ij} \label{Pro1obj1} \\
\quad & \min_{\mathbf{Y}} \sum_{(i,j) \in \mathcal{E}} T_{ij}\frac{y_{ij}}{K} \label{Pro1obj2} \\
s.t.
\quad & \sum_{i \in \mathcal{K}} x_{0i} = \sum_{i \in \mathcal{K} } x_{i0} \label{Pro1C1}\\
\quad & \sum_{i \in \mathcal{K}^a } x_{ij} = 1 , \; \forall j \in \mathcal{K} \label{Pro1C2}\\
\quad & \sum_{i \in \mathcal{K}^a } x_{ji} = 1 , \; \forall j \in \mathcal{K} \label{Pro1C3}\\
\quad & \sum_{(i,j) \in \mathcal{E} }  y_{ij} -  \sum_{(j,i) \in \mathcal{E} }  y_{ji} = 1, \; \forall i \in \mathcal{K} \label{Pro1C4}\\
\quad & y_{oi} = 0, \; \forall i \in \mathcal{K} \label{Pro1C5}\\
\quad & 0\leq y_{ij} \leq Kx_{ij}, \; \forall (i,j) \in \mathcal{E} \label{Pro1C6}\\
\quad & x_{ij} \in \{0,1\}, \; \forall (i,j) \in \mathcal{E} \label{Pro1C7}
\end{align}
\end{subequations}
\end{linenomath}
where $x_{ij}$ are binary variables and $x_{ij}=1$ represents that the edge $(i,j)$ is travelled by the UAV, $y_{ij}$ are the flow variables associated to all edges, $\mathbf{X} \triangleq \{x_{ij}\, \big| \, (i,j) \in \mathcal{E}\}$ and $\mathbf{Y} \triangleq \{y_{ij}\, \big| \, (i,j) \in \mathcal{E}\}$ are the set of variables. Observe that \eqref{Pro1C1}-\eqref{Pro1C3} impose the degree constraints for all SNs and depot. Also, \eqref{Pro1C7} reflects the binary restriction for the variable $x_{ij}$. The following two Lemmas illustrate how the constraints \eqref{Pro1C4}-\eqref{Pro1C6} operate for the aforementioned multi-return-allowed mode.

\textit{\textbf{Lemma 1:} The constraint \eqref{Pro1C4} guarantees that if a vertex set $\mathcal{K}' \subseteq \mathcal{K}^a $ constitutes a cycle, the depot must be included by $\mathcal{K}'$, i.e., $0\in \mathcal{K}'$.}

\textit{Proof:} We prove this by induction. Assume that there is a cycle without the depot, we simply choose a SN $i_1  \in \mathcal{K}'$ as the start (end) point. Thus, all SNs in this $\mathcal{K}'$ are visited in tandem and let us denote the path as $i_1 \to i_2 \to ... \to i_r \to i_1$. Now let us set $y_{i_1i_2} = c$. According to \eqref{Pro1C4}, it follows that  $y_{i_2i_3} = c + 1$, ..., $y_{i_ri_1} = c + r - 1$. Therefore, we have $\sum_{(i_1,j) \in \mathcal{E} }  y_{i_1j} - \sum_{(j,i_1) \in \mathcal{E} }  y_{ji_1} = c - (c + r - 1) = 1-r$, which contradicts with \eqref{Pro1C4}. This completes the proof of Lemma 1. $\square$

\textit{\textbf{Lemma 2:} The constraints \eqref{Pro1C4}-\eqref{Pro1C6} guarantee that the following equation is achieved,}
\begin{equation}
\label{lemma2}
\begin{aligned}
y_{ij}  = 
\begin{cases}
f_{ij}, \; &\text{if $x_{ij} = 1$} \\
0,  & \text{otherwise}
\end{cases}
\end{aligned}
\end{equation}

\textit{Proof:} Lemma 1 shows that a cycle generated by (P1) must consist of the depot. Choosing the depot as the start (end) point, the UAV would fly along the trajectory $0 \to i_1 \to i_2 ... \to i_r \to 0$. \eqref{Pro1C6} guarantees that $y_{ij}$ can achieve a nonzero value only when $x_{ij}=1$, otherwise, $y_{ij} $ is forced to 0 when $x_{ij}=0$. Recalling the constraints \eqref{Pro1C4} and \eqref{Pro1C5}, we then have $y_{0i_1} = 0, \, y_{i_1i_2} = 1, \, ..., \, y_{i_r0} = r$. Since the inequality $r \leq K$ is satisfied, the constraints \eqref{Pro1C6} would not be broken when $x_{ij} = 1$. This completes the proof of Lemma 2. $\square$

Lemma 1 illustrates that there is a feasible multi-return-allowed path generated by (P1) and Lemma 2 shows that the second objective function \eqref{Pro1obj2} calculates the average AoI. Thus, problem (P1) solves the proposed multi-return-allowed serving mode with $\overline{A}$ and $E$ as the objective functions. Also, the proposed flow-based formulation can be easily extended to characterize other AoI-aware path planning problem in aforementioned existing papers. For example, the previous paper \cite{liu2018age} limits the UAV back to the depot after all SNs been visited, which mode can be formulated as a MILP by removing the energy objective \eqref{Pro1obj1} and replacing the constraint \eqref{Pro1C1} by $\sum_{i \in \mathcal{K}} x_{0i} = 1, \; \sum_{i \in \mathcal{K} } x_{i0} = 1$. 

\section{Single Objective and Bender's decomposition}
\label{LCandBD}

Hereafter we transform (P1) to a single objective problem via the weighted linear combination technique and apply Bender's decomposition to decentralize the overall computational burden when the problem scale is large.

\subsection{Weighted Linear Combination}
\label{weightedLinear}

Various techniques are proposed to handle multi-objective optimization problems, e.g., mathematical programming methods \cite{jaimes2009introduction} and machine learning based strategies \cite{wang2020thirty}. In this paper, a widely used method is utilized, i.e., we associate the objective functions \eqref{Pro1obj1} and \eqref{Pro1obj2} with weighting coefficients and minimize the weighted sum after normalization, that is,
\begin{linenomath}
\begin{subequations}
\begin{align}
\mathrm{(P2):} 
\quad &  \min_{\mathbf{X}, \mathbf{Y}} \; \lambda \frac{\sum_{(i,j) \in \mathcal{E}} T_{ij}\frac{y_{ij}}{K} - \overline{A}_{min}}{\overline{A}_{max} - \overline{A}_{min}} \notag \\ &  \quad + (1 - \lambda) \frac{\sum_{(i,j) \in \mathcal{E}} E_{ij}x_{ij} - E_{min}}{E_{max} - E_{min}} \label{Pro2obj} \\
s.t.
\quad & \eqref{Pro1C1}-\eqref{Pro1C7} \label{Pro2C1}
\end{align}
\end{subequations}
\end{linenomath}
where $\overline{A}_{min}$ and $\overline{A}_{max}$ are the minimal and maximal achieved value of $\overline{A}$ respectively, $E_{min}$ and $E_{max}$ are the minimal and maximal energy consumption respectively, $\lambda$ and $(1-\lambda)$ denote the weights of two matrices. Theorem 4 in \cite{jaimes2009introduction} establishes that the solution of (P2) is Pareto optimal if $\lambda \in [0,1]$.

To derive an explicit expression of the objective function in (P2), the extreme values of $\overline{A}$ and $E$ should be obtained firstly. Since $\{T_{ij}\big| \, \forall (i,j) \in \mathcal{E}\}$ satisfy the triangle inequality, the UAV would return to the depot immediately after visiting each SN to minimize $\overline{A}$. The corresponding energy consumption is certainly the maximal value of $E$. Similarly, as shown in \eqref{EnergyConforEdges}, $\{E_{ij}\big| \, \forall (i,j) \in \mathcal{E}\}$ satisfy the triangle inequality and are monotonically increasing of $\lVert\mathbf{w}_i - \mathbf{w}_j\rVert$. Therefore, the most energy-efficient path would include exactly one cycle with the shortest flying distance, that is, a Travelling Salesman Problem (TSP) solution. The corresponding average AoI can be seen as $\overline{A}_{max}$. Notably, although TSP is a NP-hard problem, it has been well researched so that we assume that the TSP solution, as well as its corresponding $E_{min}$ and $\overline{A}_{max}$, are known hereafter. 

\vspace{-0.15cm}
\subsection{Applying Bender's Decomposition}
\label{BendersDecomposition}

For notational convenience, we ignore the constant terms in \eqref{Pro2obj} and rewrite (P2) as, 
\begin{linenomath}
\begin{subequations}
\begin{align}
\mathrm{(P3):} 
\quad &  \min_{\mathbf{X}, \mathbf{Y}} \; \sum_{(i,j) \in \mathcal{E}} C^T_{ij}y_{ij} + \sum_{(i,j) \in \mathcal{E}} C^E_{ij}x_{ij} \label{Pro3obj} \\
s.t.
\quad & \eqref{Pro1C1}-\eqref{Pro1C7} \label{Pro3C1}
\end{align}
\end{subequations}
\end{linenomath}
where $C^T_{ij} \triangleq \lambda T_{ij}/ (K \overline{A}_{max} - K \overline{A}_{min})$ and $C^E_{ij} \triangleq (1- \lambda) E_{ij}/ (E_{max} - E_{min})$ are coefficients defined for simplicity.

Because Bender's decomposition exploits the problem structure and decentralize the overall computation burden \cite{bnnobrs1962partitioning}, it is seen as a promising approach for large-scale MILP. To apply Bender's decomposition for (P3), we first rewrite it as the following (P4) without loss of optimality,
\begin{linenomath}
\begin{subequations}
\begin{align}
\mathrm{(P4):} 
\quad &  \min_{\mathbf{X}} \; \sum_{(i,j) \in \mathcal{E}} C^E_{ij}x_{ij} + g(\mathbf{X}) \label{Pro4obj} \\
s.t.
\quad & \eqref{Pro1C1}-\eqref{Pro1C3}, \;\eqref{Pro1C7} \label{Pro4C1}
\end{align}
\end{subequations}
where $g(\mathbf{X})$ is defined to be the optimal solution of the following problem,
\begin{subequations}
\begin{align}
\mathrm{(P5):} 
\quad &  \min_{\mathbf{Y}} \sum_{(i,j) \in \mathcal{E}} C^T_{ij}y_{ij} \label{Pro5obj} \\
s.t.
\quad & \eqref{Pro1C4}-\eqref{Pro1C6} \label{Pro5C1}
\end{align}
\end{subequations}
\end{linenomath}
(P5) is certainly a linear programming with respect to $\mathbf{Y}$ for given value of $x_{ij} \in \mathbf{X}$. Then, write the dual for (P5) as,
\begin{linenomath}
\begin{subequations}
\begin{align}
& \mathrm{(P6):} 
\;  \max_{ \{\alpha_i\} \{\beta_i\} \{\gamma_{ij} \} } \sum_{i \in \mathcal{K}} \alpha_i - \sum_{(i,j) \in \mathcal{E}} K x_{ij} \gamma_{ij} \label{Pro6obj} \\
& s.t.
\; -\alpha_i + \beta_i - \gamma_{0i} \leq C^T_{0i}, \; \forall i \in \mathcal{K}  \label{Pro6C1} \\
& \quad \quad \alpha_i - \gamma_{i0} \leq C^T_{i0}, \; \forall i \in \mathcal{K}  \label{Pro6C2} \\
& \quad \quad\alpha_i - \alpha_j - \gamma_{ij} \leq C^T_{ij}, \; \forall (i,j) \in \mathcal{E}, \, i \neq 0, \, j \neq 0  \label{Pro6C3} \\
& \quad \quad \gamma_{ij} \geq 0, \; \forall (i,j) \in \mathcal{E}, \label{Pro6C4}
\end{align}
\end{subequations}
\end{linenomath}
where $\{\alpha_i \,\big|\, i \in \mathcal{K}\}$, $\{\beta_i \,\big|\, i \in \mathcal{K}\}$ and $\{\gamma_{ij} \,\big|\, (i,j) \in \mathcal{E}\}$ are dual variables. The key observation is that the constraints \eqref{Pro6C1}-\eqref{Pro6C4} are not related to the values of $\mathbf{X}$. Supposing the polyhedron constructed by the constraints \eqref{Pro6C1}-\eqref{Pro6C4} has $M$ extreme points and $N$ extreme rays, (P6) has an equivalent form as follows \cite{bnnobrs1962partitioning}.
\begin{linenomath}
\begin{subequations}
\begin{align}
& \mathrm{(P7):} 
\quad   \min_{\theta} \theta \label{Pro6obj} \\
& s.t.
\; \sum_{i \in \mathcal{K}} \alpha^{p,m}_i - \! \! \sum_{(i,j) \in \mathcal{E}} K x_{ij} \gamma^{p,m}_{ij} \leq \theta, \; \forall m = 1,...,M  \label{Pro7C1} \\
& \quad \; \sum_{i \in \mathcal{K}} \alpha^{r,n}_i - \! \sum_{(i,j) \in \mathcal{E}} K x_{ij} \gamma^{r,n}_{ij} \leq 0, \; \forall n = 1,...,N  \label{Pro7C2}
\end{align}
\end{subequations}
\end{linenomath}
where $(\alpha^{p,m}_1,...,\alpha^{p,m}_K, \beta^{p,m}_1,...,\beta^{p,m}_K,\gamma^{p,m}_{01},...,\gamma^{p,m}_{K-1K})$ and $(\alpha^{r,n}_1,...,\alpha^{r,n}_K, \beta^{r,n}_1,...,\beta^{r,n}_K,\gamma^{r,n}_{01},...,\gamma^{r,n}_{K-1K})$ denote the $m^{th}$ extreme point and $n^{th}$ extreme ray of the polyhedra constructed by \eqref{Pro6C1}-\eqref{Pro6C4}, respectively. And the constraints in \eqref{Pro7C1} and \eqref{Pro7C2} are named as optimality and feasibility cuts, respectively. Since (P5) is a linear programming of which the strong duality is held, (P5), (P6) and (P7) achieve the same optimal solution. Accordingly, replacing the component $g(\mathbf{X})$ by (P7), (P4) can be rewritten as the following form without loss of optimality,
\begin{linenomath}
\begin{subequations}
\begin{align}
\mathrm{(P8):} 
\quad &  \min_{\mathbf{X},\theta} \; \sum_{(i,j) \in \mathcal{E}} C^E_{ij}x_{ij} + \theta \label{Pro7obj} \\
s.t.
\quad & \eqref{Pro1C1}-\eqref{Pro1C3}, \;\eqref{Pro1C7} \label{Pro8C1} \\
\quad & \eqref{Pro7C1}-\eqref{Pro7C2} \label{Pro8C2} 
\end{align}
\end{subequations}
\end{linenomath}
(P8) and (P5) are always named as master problem and sub-problem, respectively.

\begin{algorithm}[!t]
\small
\caption{Applying Bender's Decomposition to (P3)}
\label{BDalg}
\begin{algorithmic}[1]
\STATE Initialize the RMP without any cuts in \eqref{Pro8C2}. Set $k = 0$.
\REPEAT
\STATE Solve the RMP. Denote the solution as $\mathbf{X}^k$ and $\theta^k$. \label{Alg1step2}
\STATE Introduce $\mathbf{X}^k$ to the dual sub-problem (P6). Solve (P6) and denote the optimal objective value as $g(\mathbf{X}^k)$. \label{Alg1step3}
\IF{$g(\mathbf{X}^k) < \infty$} 
\STATE Acquire an extreme point from the solution of (P6)  and add an optimality cut \eqref{Pro7C1} to RMP.
\ELSE
\STATE Acquire an extreme ray and add a feasibility cut \eqref{Pro7C2}.
\ENDIF
\STATE $k = k+1$.
\UNTIL{$g(\mathbf{X}^k) = \theta^k$} \label{Alg1step11}
\end{algorithmic}
\end{algorithm}

However, since the number of the constraints in \eqref{Pro7C1} and \eqref{Pro7C2} is extremely large, generating all of them is impractical. To this end, we construct a problem having a similar form as (P8) but which does not consist of all the constraints in \eqref{Pro8C2} as the relaxed master problem (RMP). According to the nominal use of Bender's decomposition, adding the optimality and feasibility cuts according to the solution of dual sub-problem (P6) to the RMP and solving it iteratively. The procedure of Bender's decomposition is summarized as Algorithm \ref{BDalg}, a realization of which implemented by the Python API of Gurobi 9.1.2 \cite{gurobi} can be found in \href{https://github.com/Yuanliaoo/BDforUavPath}{github.com/Yuanliaoo/BDforUavPath}.

We then review the optimality and convergence of Bender's decomposition. Firstly, denoting the optimal solution of (P3) as $\psi^*$, section 4 in \cite{bnnobrs1962partitioning} shows that at iteration $k$, the lower and upper bounds of $\psi^*$ can be obtained by step \ref{Alg1step2} and \ref{Alg1step3} of Algorithm \ref{BDalg} by $LB^k = \sum_{(i,j) \in \mathcal{E}} C^E_{ij}x_{ij}^k + \theta^k$ and $UB^k = \sum_{(i,j) \in \mathcal{E}} C^E_{ij}x_{ij}^k + g(\mathbf{X}^k)$, respectively. i.e. $LB^k \leq \psi^* \leq UB^k$. Therefore, when the gap between $LB^k$ and $UB^k$ is equal to 0, the global optimal solution $\psi^*$ is surely achieved, shown as the stopping criteria in the step \ref{Alg1step11} of Algorithm \ref{BDalg}. Also, Theorem 3.1 and section 4 in \cite{bnnobrs1962partitioning} illustrate that the procedures would terminate and converge to $\psi^*$ within finite number of iterations.

\section{General Trajectory Optimization}
\label{FurtherOpt}

The fly-hover-communication protocol in the preceding section determines the visiting order of the UAV. However, both the average AoI and energy can be further optimized when the UAV is allowed to collect data while flying instead of the binary flying status. In this section, assuming the visiting order has been solved by (P1), the UAV trajectory is further optimized to improve the system performances without the fly-hover-communication assumption.

\begin{figure}[!t]
\centering
\subfigure{\includegraphics[width=.38\textwidth]{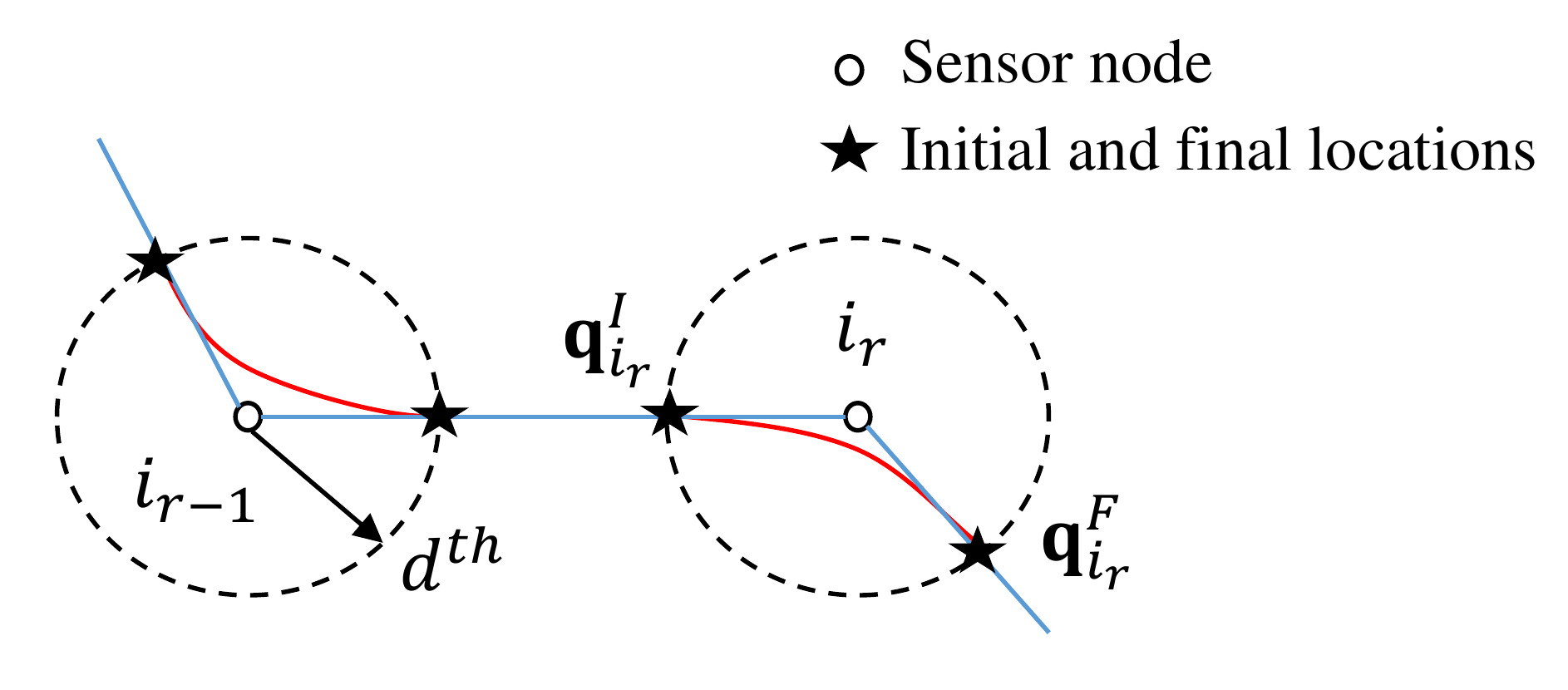}}
\caption{Illustration of further trajectory optimization. The blue line represents the path solved by (P1) while the red line shows a possible solution of the optimized trajectory obtained by (P9).}
\label{Trajectory_opt}
\end{figure}

Denote the UAV location at time $t$ projected onto the horizontal plane by $\mathbf{q}(t) \in  \mathbb{R}^2$. The achievable rate between UAV and the SN $i$  at time $t$ can be rewritten from \eqref{AchRate} as a function of $\mathbf{q}(t)$,
\begin{equation}
\label{ReAchRate}
\begin{aligned}
R(\mathbf{q}(t)) = B \cdot \log_2 \Big( 1 + \frac{ P^{t} \rho_0}{\sigma^2(H^2+\lVert \mathbf{q}(t) - \mathbf{w}_i\rVert^2) }\Big)
\end{aligned} 
\end{equation}
To guarantee the successful decoding and the quality of service (QoS), the signal-to-noise-ratio (SNR) at the UAV, defined by SNR $\triangleq \frac{ P^{t} \rho_0}{\sigma^2(H^2+\lVert \mathbf{q}(t) - \mathbf{w}_i\rVert^2) }$, is required to greater than a pre-specified threshold \cite{mozaffari2016mobile,alzenad20173}. Because SNR is monotonically decreasing with $\lVert \mathbf{q}(t) - \mathbf{w}_i\rVert^2$, this QoS requirement can be satisfied by a distance constraint, i.e., $\lVert \mathbf{q}(t) - \mathbf{w}_i\rVert^2 \leq d^{th}$. As shown in Fig.\ref{Trajectory_opt}, the UAV can only receive data from SN $i$ when it is located in a circular disc with the center $\mathbf{w}_i$ and radius $d^{th}$, which is named as coverage area hereafter.

The UAV trajectory only needs to be determined within the coverage areas because in other area the UAV would fly along line segments to decrease both the flying time and energy. Firstly, we would select the initial and final locations for the UAV trajectory within a certain circular coverage area. As shown in Fig.\ref{Trajectory_opt}, when the UAV visits $i_{r-1}$ and $i_r$ adjacently, the intersection point of the line segment $\mathbf{w}_{i_{r-1}}\mathbf{w}_{i_{r}}$ and the edge of the coverage area is defined as the initial location of the UAV trajectory when it flies and connects to SN $i_{r}$, denoted by $\mathbf{q}^I_{i_r}$, while the intersection of $\mathbf{w}_{i_{r}}\mathbf{w}_{i_{r+1}}$ and the edge is defined as the final point and denoted by $\mathbf{q}^F_{i_{r}}$. Once the start and final locations are fixed, the UAV trajectory when flying across the coverage area of SN $i_r$ can be optimized by the following problem,
\begin{linenomath}
\begin{subequations}
\begin{align}
\mathrm{(P9):} 
\quad &  \min_{\{\mathbf{q}(t)\}, \, T_{i_r}} \int_0^{T_{i_r}} P^{f}(\lVert \dot{\mathbf{q}}(t) \rVert) \; dt \label{Pro9obj1} \\
\quad & \min_{\{\mathbf{q}(t)\}, \, T_{i_r}}  \frac{f_{i_{r-1}i_r}}{K} \, T_{i_r} \label{Pro9obj2} \\
s.t.
\quad & \int_0^{T_{i_r}}R(\mathbf{q}(t)) \, dt \geq D_{i_r} \label{Pro9C1}\\
\quad & \mathbf{q}(0) = \mathbf{q}^I_{i_r}, \quad \mathbf{q}(T_{i_r}) = \mathbf{q}^I_{i_r} \label{Pro9C2}\\
\quad & \lVert \mathbf{q}(t) - \mathbf{w}_{i_r}\rVert^2 \leq d^{th}, \; \forall t \in [0,T_{i_r}] \label{Pro9C3}\\
\quad & \lVert \dot{\mathbf{q}}(t) \rVert \leq V^{max}, \; \forall t \in [0,T_{i_r}] \label{Pro9C4}
\end{align}
\end{subequations}
\end{linenomath}
where $T_{i_r}$ denotes the flying time when the UAV travels through the coverage area of SN $i_r$, $\lVert \dot{\mathbf{q}}(t) \rVert$ calculates the norm of the UAV velocity at time $t$ and $V^{max}$ denotes the maximum UAV speed. The objective functions \eqref{Pro9obj1} and \eqref{Pro9obj2} optimize the partial energy consumption and average AoI, respectively. \eqref{Pro9C1} guarantees that the data collection is completed during this period and \eqref{Pro9C4} is the maximum speed constraint.

Problem (P9) is challenging to solve for the following two main reasons. First, (P9) involves an infinite number of variables $\{\mathbf{q}(t)\}$ that is over continuous time $t$. Second, both the objective function \eqref{Pro9obj1} and the constraint \eqref{Pro9C1} are non-convex with respect to decision variables. Fortunately, after being converted to a single objective problem by linear combination technique, (P9) has been solved perfectly in the previous work \cite{zeng2019energy}. Specifically, in the section IV of \cite{zeng2019energy}, the path discretization technique is applied to disretize the variables $\{\mathbf{q}(t)\}$ and the successive convex approximation (SCA) method is then utilized to solve the non-convex problem. More details can be found in \cite{zeng2019energy} and a realization coded by MATLAB and CVX \cite{cvx} can be found in \href{https://github.com/Yuanliaoo/AoIEnergyUAVTraOpt}{github.com/Yuanliaoo/AoIEnergyUAVTraOpt}.

\section{Numerical Investigations}
\label{NumericalResults}

\begin{table}[!t]
\centering
\caption{Parameter Settings}
\label{TAB para}
\begin{tabular}{ll|ll}
\hline
Parameter & Value & Parameter & Value\\
\hline
$B$ & 2 MHz & $\sigma^2$ & -110 dBm\\
$H$ & 100 m & $\rho_0$ & -60 dB \\
$P^t$ & 0.1 W & $D_i$ & 500 Mbits \\
$K$ & 10 (expect Fig.\ref{Comparing}) & $V$ & [10, 18, 30] m/s \cite{zeng2019energy} \\
$P^h$ & 165 W \cite{zeng2019energy}& $P^f$ & [126, 162, 356] W \cite{zeng2019energy} \\
$d^{th}$ & 50 m & $V^{max}$ & 30 m/s \cite{zeng2019energy} \\
\hline
\end{tabular}
\vspace{-0.2cm}
\end{table}

In this section, numerical investigations are presented to evaluate the proposed multi-return-allowed framework. The parameter settings are summarized in Table \ref{TAB para}. We mainly compare the metrics under the following three typical choices of speed, \textit{Maximum-endurance (ME) speed} maximizes the UAV endurance under any given on-board energy; \textit{Maximum-range (MR) speed} maximizes the total traveling distance with any given on-board energy; \textit{Maximal (MAX) speed} is the maximal achievable speed of the UAV. As illustrated in \cite{zeng2019energy}, the ME, MR and MAX speeds are 10 m/s, 18 m/s and 30 m/s, respectively. And the corresponding values of the propulsion power are 126 W, 162 W and 356 W, as shown in Table \ref{TAB para}.

\begin{figure}[!t]
\centering
\subfigure{\includegraphics[width=.42\textwidth]{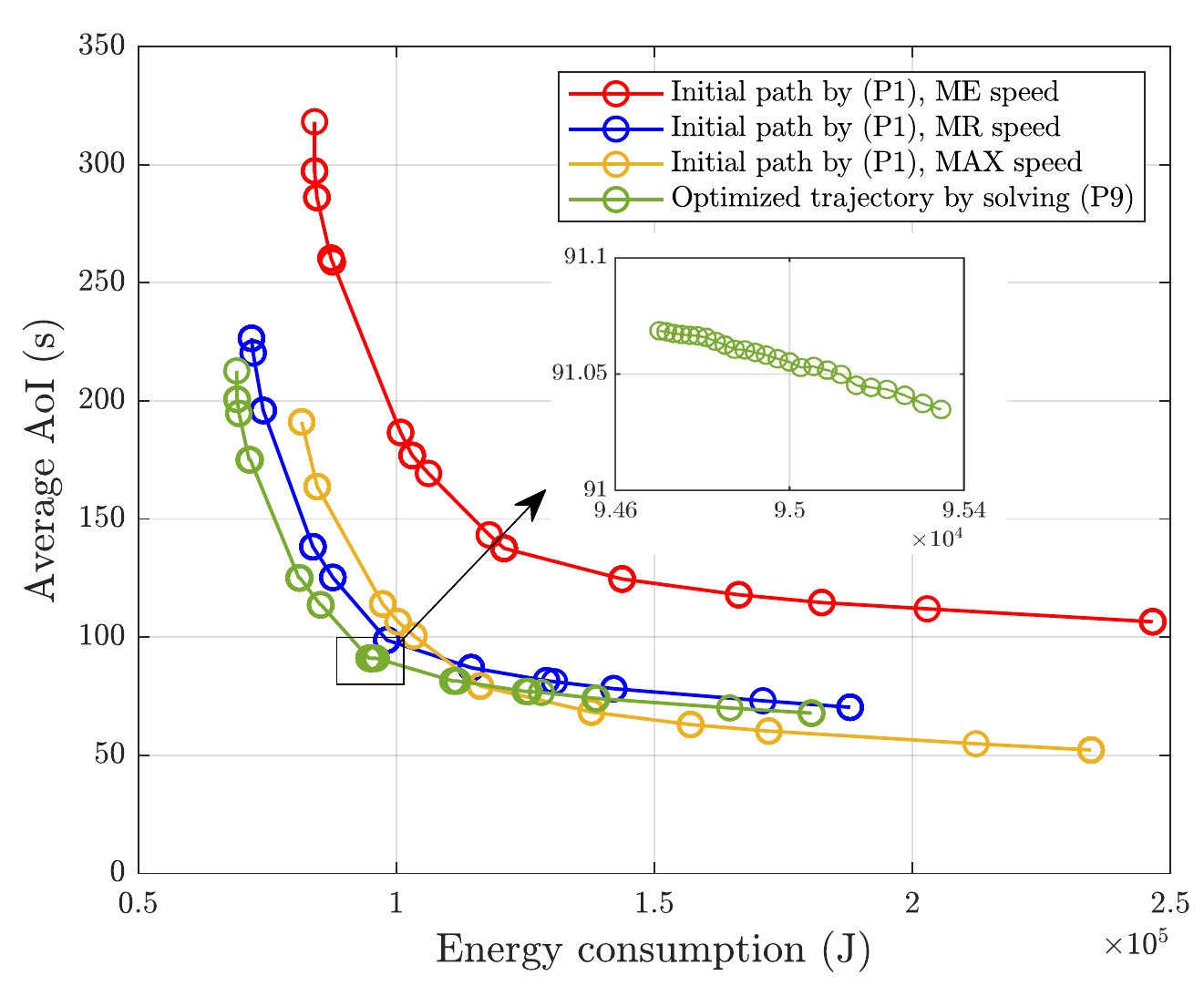}}
\caption{Pareto front for $\overline{A}$ and $E$ with different speed}
\label{Pareto_front}
\end{figure}

Fig.\ref{Pareto_front} depicts the Pareto front of $\overline{A}$ and $E$. Firstly, solving (P2) for different values of $\lambda$, the obtained results, represented by red, yellow and blue small circles, are in accordance with the intuition that $\overline{A}$ and $E$ are competing with each other. It is worth pointing out that although we sample the value of $\lambda \in [0,1]$ every 0.01 to plot Fig.\ref{Pareto_front}, the actual results are distributed sparsely and overlap. The reason is that both $\overline{A}$ and $E$ are functions of the UAV path, which are determined by the binary variables $\mathbf{X}$ in (P1). 
Furthermore, optimizing the initial path with MR speed by solving (P9), the obtained result, represented by green circles in Fig.\ref{Pareto_front}, shows the improvement in both the $\overline{A}$ and $E$ whatever the value of $\lambda$ is. Besides, these Pareto points are distributed more densely because the variables in (P9) can take continuous values. 

\begin{figure}[!t]
\centering
\subfigure{\includegraphics[width=.44\textwidth]{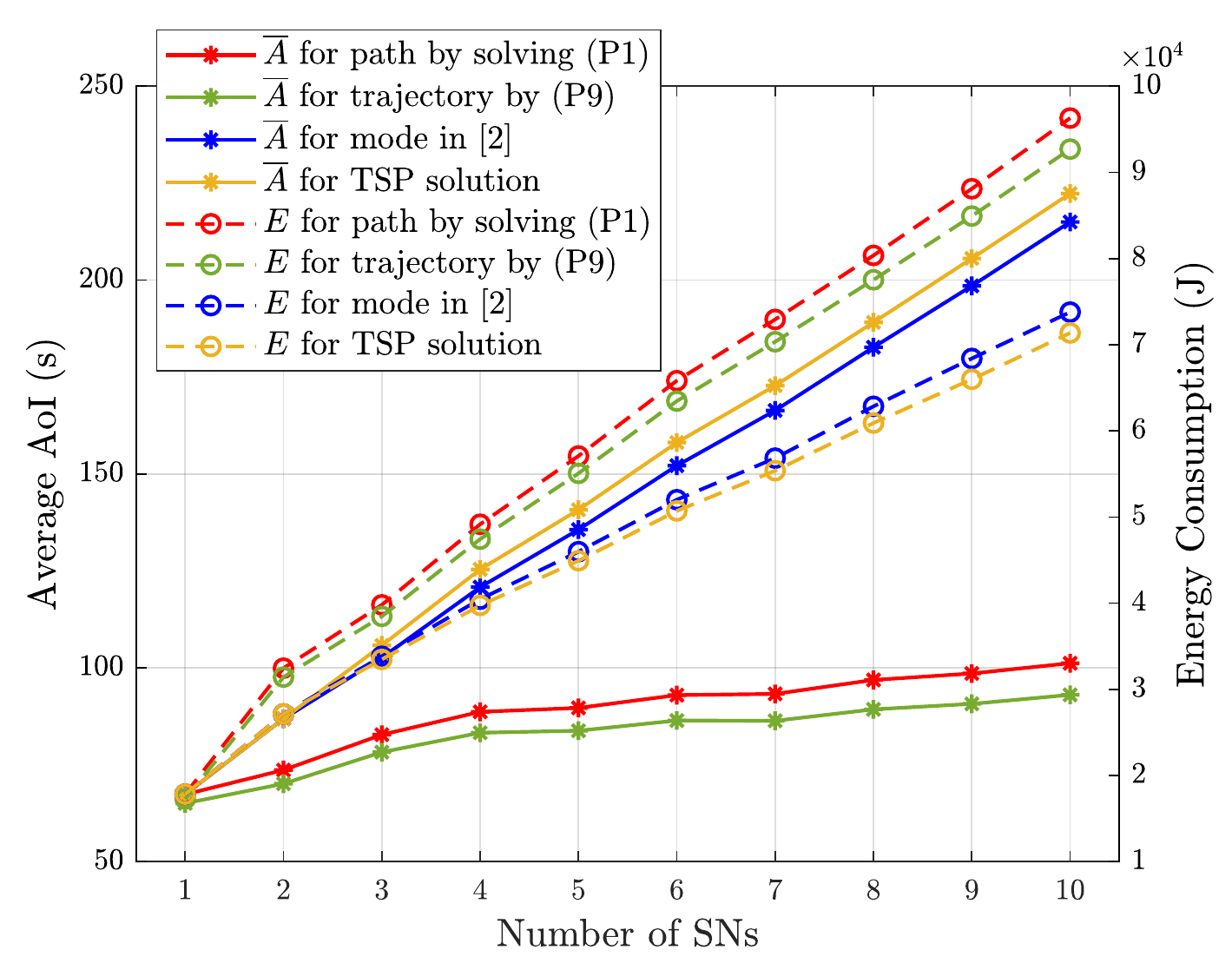}}
\caption{Comparing $\overline{A}$ and $E$ for different serving modes}
\label{Comparing}
\end{figure}

In Fig.\ref{Comparing}, we compare the proposed multi-return-allowed mode (setting $\lambda = 0.5$) with the framework studied in \cite{liu2018age}, in which the UAV returns to the depot only after all SNs been visited, as well as the TSP solution. Firstly, comparing the three serving mode under the fly-hover-communication assumption, denoted by red, blue and yellow lines, it can be observed that the proposed multi-return-allowed mode can decrease $\overline{A}$ significantly at the cost of a larger $E$. Numerically, take the $K=10$ as an example, the mode in \cite{liu2018age} decreases $\overline{A}$ by 3\% at the cost of 3\% more energy consumption when comparing with the TSP solution. Furthermore, the proposed framework shows a 52\% gain in $\overline{A}$ than mode \cite{liu2018age} with a 29\% higher energy consumption. Secondly, comparing the path initialized by (P1) and the trajectory optimized by (P9), it can be seen that both the $\overline{A}$ and $E$ are further improved when the UAV can communicate while flying. For instance, $\overline{A}$ and $E$ decrease by 8\% and 4\% respectively in the optimized trajectory when $K=10$. 

\section{Conclusions}
\label{Conclusion}

This paper studies the path planning problem of a UAV-assisted data collection task, in which two inherently competing metrics, namely the average AoI and the aggregate energy consumption, are optimized jointly. To characterize the trade-off between those two performance metrics and reveal the Pareto frontier, we propose a multi-return-allowed serving mode in which the UAV is allowed to return to the depot at any instance during the service cycle, and formulate it as a multi-objective MILP with a flow-based constraint set. The two objectives are combined into a single one through the weighted linear combination technique, and we apply the Bender's decomposition on the reformulated MILP to decentralize the overall computational burden. Subsequently, a more general trajectory in which the UAV can communicate while flying is studied to further improve both these metrics. Additionally, previous research works that solely focus on Hamiltonian paths can be considered as a special case of the proposed approach since it allows multiple returns to the depot during a serving cycle. A wide set of numerical investigations reveal that the proposed multi-return-allowed mode unveils the trade-off between the two competing metrics and hence, provides non-dominated solutions for advanced decision making. 

\bibliographystyle{IEEEtran}
\bibliography{IEEEabrv,reference} 

\end{document}